\documentclass[pra,aps,twocolumn,showpacs]{revtex4-1}
\usepackage{amssymb}
\usepackage{hyperref}
\usepackage{graphicx}
\usepackage{dcolumn}
\usepackage{bm,amsmath,verbatim}

\def\be{\begin{equation}}
\def\ee{\end{equation}}
\def\bea{\begin{eqnarray}}
\def\eea{\end{eqnarray}}
\def\bse{\begin{subequations}}
\def\ese{\end{subequations}}

\def\be{\begin{eqnarray}}
\def\ee{\end{eqnarray}}

\begin{document}

\title{Competing superfluid orders in spin-orbit coupled fermionic cold atom
optical lattices}
\author{Yong Xu$^{1}$}
\author{Chunlei Qu$^{1}$}
\author{Ming Gong$^{1,2}$}
\author{Chuanwei Zhang$^{1}$}
\thanks{Corresponding Author, Email: chuanwei.zhang@utdallas.edu}

\begin{abstract}
The Fulde-Ferrell-Larkin-Ovchinnikov (FFLO) phase, a superconducting state
with non-zero total momentum Cooper pairs in a large magnetic field, was
first predicted about 50 years ago, and since then became an important
concept in many branches of physics. Despite intensive search in various
materials, unambiguous experimental evidence for the FFLO phase is still
lacking in experiments. In this paper, we show that both FF (uniform order
parameter with plane-wave phase) and LO phase (spatially varying order
parameter amplitude) can be observed using fermionic cold atoms in
spin-orbit coupled optical lattices. The increasing spin-orbit coupling
enhances the FF phase over the LO phase. The coexistence of superfluid and
magnetic orders is also found in the normal BCS phase. The pairing mechanism
for different phases is understood by visualizing superfluid pairing
densities in different spin-orbit bands. The possibility of observing
similar physics using spin-orbit coupled superconducting ultra-thin films is
also discussed.
\end{abstract}
\affiliation{$^{1}$Department of Physics, The University of Texas at Dallas, Richardson,
Texas, 75080 USA \\
$^{2}$ Department of Physics, The Chinese University of Hong Kong, Hong Kong}
\maketitle

\section{Introduction}

The interplay between magnetism and superconductivity leads to various
interesting phenomena, which have been intensively studied in many\
different materials~\cite{FuldePR1964,Larkin1964,PickettPRL1999,LynnPRB2000,
SaxenaNature2000,KenzelmannScience2008,Li2011Nature,BertNature2011,DikinPRL2011}%
. The physics from such interplay can become even richer and more important
when there exists strong spin-orbit (SO) coupling \cite%
{Xiao2010RMP,Kane2010RMP,Zutic2004RMP} in underlying physical systems, as
evidenced by the recent impressive progress on the search for Majorana
fermions using superconductor-semiconductor nanowire heterostructures~\cite%
{MourikScience2012,DengMTNano2012,DasNature2012,LPRnature2012}. Another
well-known physics originating from the interplay between magnetism and
superconductivity is the Fulde-Ferrell-Larkin-Ovchinnikov~(FFLO)
superconducting phase~\cite{FuldePR1964,Larkin1964}, where electrons form
Cooper pairings with non-zero center-of-mass momentum in the presence of a
large Zeeman field. In the past five decades, intensive experimental search
for FFLO phases has been done in different materials \cite%
{HeaFermFFLO,KenzelmannScience2008,OrgSupFFLO,Cho2011PRB,Li2011Nature,Liao2010Nature}%
. However, unambiguous experimental evidences for FFLO states are still
elusive in experiments.

Ultracold degenerate Fermi gases may provide an ideal platform for exploring
FFLO physics because of their intrinsic advantages such as the lack of
orbital effects, free of disorder, as well as highly controllable
experimental parameters. While previously FFLO phases have been widely
studied in spin-imbalanced Fermi gases \cite{FFLOFreeGas}, the recent
experimental realization of SO coupling for cold atoms \cite{SOexp} provides
a completely new route for the experimental observation of FFLO phases. Note
that there are two different types of FFLO phases: FF (Fulde-Ferrell) state~%
\cite{FuldePR1964} with uniform amplitude but spatially dependent phase, and
LO (Larkin-Ovchinnikov) \cite{Larkin1964} state with spatially oscillating
amplitude but uniform phase of the order parameter.

In this paper, we show that both FF phase and a generalized LO phase may be
observed in SO coupled fermionic cold atom optical lattices. Here the
superfluid order parameters are obtained by self-consistently solving the
real space Bogoliubov-de-Gennes (BdG) equation. Without SO coupling, it is
well known that LO states emerge in lattices with a large Zeeman field \cite%
{Trivedi2010PRL}. With SO coupling, FF phases have been proposed in free
space without lattices \cite{FreeGasSOC,Barzykin2002PRL,Partrick2012PRL},
where the superfluid gap equation is solved in the momentum space (thus the
spatial oscillation of the LO phase cannot be found). Our real space BdG
equation can capture both FF and LO phases, and we show that there is a
competition between them in a SO coupled optical lattice \cite%
{chunlei2013,IskinArXiv}. Generally SO coupling enhances the FF phase while
suppresses the LO phase. The generalized LO phase has no spatially nodes in
the order parameter and magnetization, which are very different from
traditional LO states in spin-imbalanced Fermi gases. The BCS pairing order
also possesses finite magnetization, showing that the coexistence of
magnetism and superfluid does not necessarily indicate the existence of FFLO
phases \cite{Liao2010Nature}. The pairing mechanism for FF and LO phases is
understood by visualizing the superfluid pairing densities in different SO
bands. Finally, we discuss the possibility of observing similar physics
using SO coupled superconducting ultra-thin films (e.g., Pb) \cite%
{Qikun2010Nature,Slomski2013,Xiong2011Natrue}.

\section{Theoretical model}

We consider fermionic cold atoms confined in a two-dimensional (2D) Rashba
SO coupled square optical lattice and subject to an in-plane Zeeman field.
The system can be described by a Fermi-Hubbard Hamiltonian%
\begin{equation}
H=H_{0}+H_{SO}+H_{Z},  \label{tb_ham}
\end{equation}%
where
\begin{equation}
H_{0}=-t\sum_{\langle ij\rangle \sigma }\hat{c}_{i\sigma }^{\dagger }\hat{c}%
_{j\sigma }-\mu \sum_{i\sigma }\hat{n}_{i\sigma }+U\sum_{i}\hat{n}%
_{i\uparrow }\hat{n}_{i\downarrow }
\end{equation}
is the usual single particle Hamiltonian with an on-site interaction, $\hat{c%
}_{i\sigma }$ is the atom annihilation operator at the $i$-th site with spin
$\sigma $, and $\hat{n}_{i\sigma }$ is the particle number operator. $t$, $%
\mu $, and $U$ represent hopping strength, chemical potential, and on-site
interaction strength, respectively.
\begin{equation}
H_{SO}=-i\alpha \sum_{\langle ij\rangle }\hat{c}_{i}^{\dagger }(\mathrm{%
\mathbf{d}}_{ij}\times \hat{\sigma}\cdot \mathrm{\mathbf{e}}_{z})\hat{c}_{j}
\end{equation}
is the Rashba SO coupling with $\hat{c}_{i}=\left(
\begin{array}{cc}
\hat{c}_{i\uparrow } & \hat{c}_{i\downarrow }%
\end{array}%
\right) $, the Pauli matrix $\hat{\sigma}$, and the unit bond vector $%
\mathrm{\mathbf{d}}_{ij}$ between nearest-neighbor sites $i$ and $j$.
\begin{equation}
H_{Z}=h\sum_{i}(\hat{c}_{i\uparrow }^{\dagger }\hat{c}_{i\downarrow }+\hat{c}%
_{i\downarrow }^{\dagger }\hat{c}_{i\uparrow })
\end{equation}
is an in-plane Zeeman field.

The superfluid phases can be studied under the standard mean-field
approximation, where the on-site interaction can be decomposed as%
\begin{equation}
U\hat{n}_{i\uparrow }\hat{n}_{i\downarrow }\approx \Delta _{i}^{\ast }\hat{c}%
_{i\downarrow }\hat{c}_{i\uparrow }+\Delta _{i}\hat{c}_{i\uparrow }^{\dagger
}\hat{c}_{i\downarrow }^{\dagger }-|\Delta _{i}|^{2}/U+\text{HFC.}
\end{equation}
Here the order parameter $\Delta _{i}=U\langle \hat{c}_{i\downarrow }\hat{c}%
_{i\uparrow }\rangle $, and the Hartree-Fock correction (HFC) terms are $%
|n_{ix}|^{2}/U-n_{ix}\hat{c}_{i\downarrow }^{\dagger }\hat{c}_{i\uparrow
}+n_{ix}^{\ast }\hat{c}_{i\uparrow }^{\dagger }\hat{c}_{i\downarrow
}+U\langle \hat{n}_{i\uparrow }\rangle \hat{n}_{i\downarrow }+U\langle \hat{n%
}_{i\downarrow }\rangle \hat{n}_{i\uparrow }-U\langle \hat{n}_{i\uparrow
}\rangle \langle \hat{n}_{i\downarrow }\rangle $ with $n_{ix}=U\langle \hat{c%
}_{i\uparrow }^{\dagger }\hat{c}_{i\downarrow }\rangle $. The mean-field
Hamiltonian can be diagonalized by the Bogoliubov transformation, $\hat{c}%
_{i\sigma }=\sum_{n=1}^{n=2N}\left( u_{i\sigma }^{n}\hat{\gamma}_{n}-{\sigma
}v_{i\sigma }^{n\ast }\hat{\gamma}_{n}^{\dagger }\right) $ with
quasi-particle operators $\hat{\gamma}_{n}$ and $\hat{\gamma}_{n}^{\dagger }$%
, yielding the BdG equation%
\begin{equation}
\sum\nolimits_{j}\left(
\begin{array}{cc}
H_{ij} & \Delta _{ij} \\
\Delta _{ij}^{\ast } & -\sigma _{y}H_{ij}^{\ast }\sigma _{y}%
\end{array}%
\right) \Phi _{j}^{n}=E_{n}\Phi _{i}^{n}.  \label{BdG}
\end{equation}%
Here $H_{ij}$ is a $2\times 2$ matrix with components $H_{ij}(\sigma \sigma
)=-t\delta _{\langle ij\rangle }-\delta _{ij}\tilde{\mu}_{i\sigma }$, $%
H_{ij}(\uparrow \downarrow )=(h-n_{ix}^{\ast })\delta _{ij}+i\alpha (\mathrm{%
\mathbf{d}}_{ij}\times \hat{\sigma}\cdot \mathrm{\mathbf{e}}_{z})_{12}\delta
_{\langle ij\rangle }$, and $H(\downarrow \uparrow )=H(\uparrow \downarrow
)^{\ast }$. $\delta _{\langle ij\rangle }=1$ for nearest neighbors, zero
otherwise; $\tilde{\mu}_{i\sigma }=\mu -U\langle \hat{n}_{i\overline{\sigma }%
}\rangle $. The quasiparticle wavefunction $\Phi _{j}^{n}=\left(
u_{j\uparrow }^{n},u_{j\downarrow }^{n},v_{j\downarrow }^{n},v_{j\uparrow
}^{n}\right) ^{T}$. The BdG equation should be solved self-consistently with
the atom density and pairing order parameter equations
\begin{eqnarray}
&&\langle \hat{n}_{i\sigma }\rangle =\sum_{n=1}^{2N}\left[ |u_{i\sigma
}|^{2}f(E_{n})+|v_{i\sigma }|^{2}f(-E_{n})\right] , \\
&&n_{ix} =-U\sum_{n=1}^{2N}\left[ v_{i\uparrow }^{n}v_{i\downarrow }^{n\ast
}(1-f(E_{n}))-u_{i\uparrow }^{n\ast }u_{i\downarrow }^{n}f(E_{n})\right] , \\
&&\Delta _{ij} =-U\delta _{ij}\sum_{n=1}^{2N}\left[ u_{i\uparrow
}^{n}v_{i\downarrow }^{n\ast }(1-f(E_{n}))-v_{i\uparrow }^{n\ast
}u_{i\downarrow }^{n}f(E_{n})\right].
\end{eqnarray}
Here the Fermi-Dirac distribution $f(E_{n})=1/\left( 1+\exp \left(
E_{n}/k_{B}T\right) \right) $. The ground state energy is given by%
\begin{eqnarray}
\langle H\rangle &=&\sum_{n=1}^{2N}E_{n}\left[ f(E_{n})-\sum_{i\sigma
}|v_{i\sigma }^{n}|^{2}\right]  \notag \\
&&+\sum_{i}\left( |\Delta _{i}|^{2}/U+U\langle n_{i\uparrow }\rangle \langle
n_{i\downarrow }\rangle -|n_{ix}|^{2}/U\right) .
\end{eqnarray}

In our numerical simulation, we choose different initial inputs of $\Delta
_{i}$ and self-consistently solve the BdG Eq. (\ref{BdG}) with a periodic
boundary condition to calculate superfluid order parameters. If different
phases are obtained for different initial inputs, we compare the energies of
these phases to determine the ground state.

\begin{figure}[t]
\includegraphics[width=3.4in]{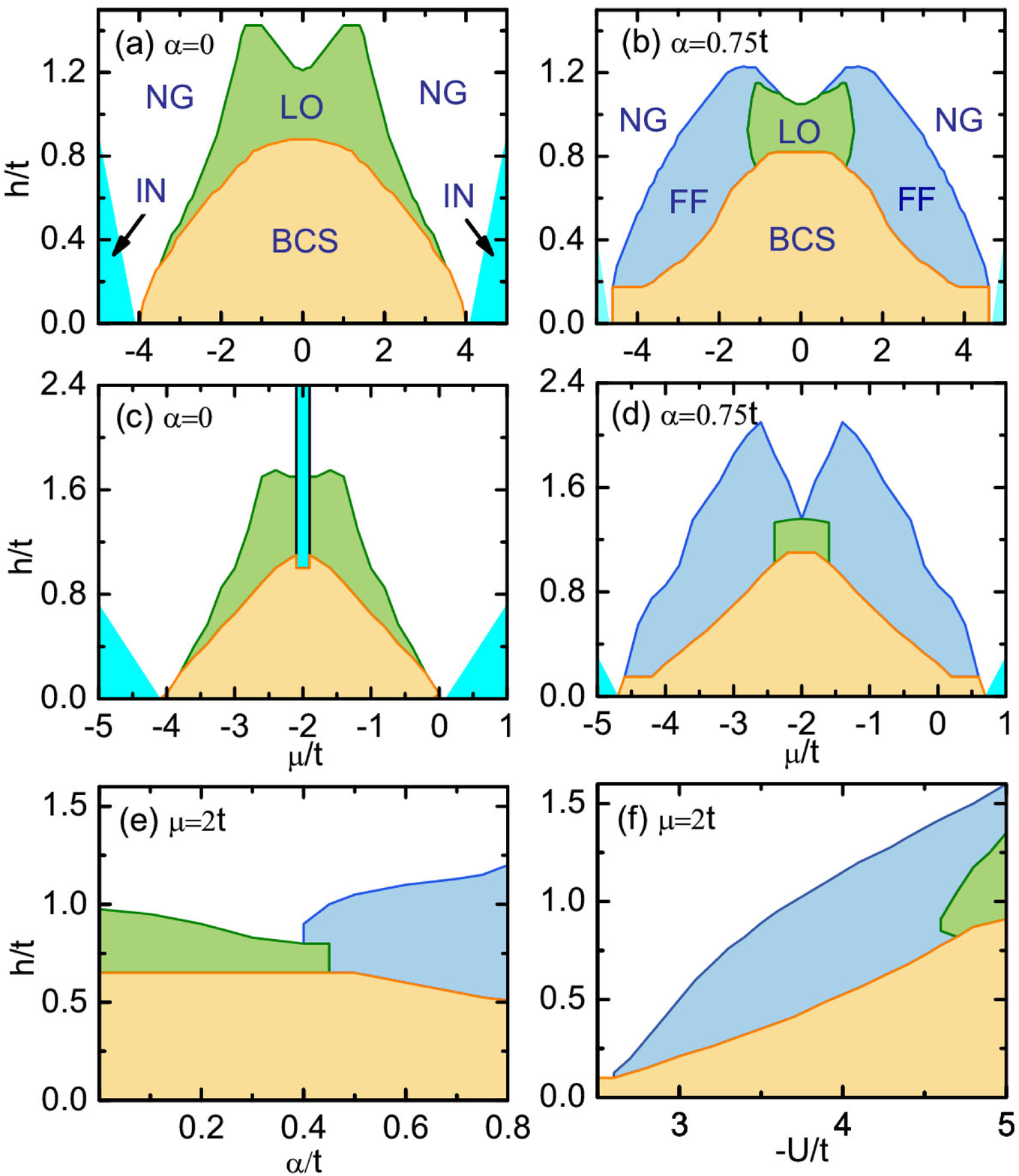}
\caption{(Color online) Mean-field phase diagrams as a function of chemical
potential $\protect\mu $ and Zeeman field $h$, in the absence of HFC for (a)
and (b), and in the presence of HFC for (c) and (d). (e) and (f) are phase
diagrams (without HFC) with varying $\protect\alpha$ and $U$ at $\protect\mu%
=2t$. We use a $16\times 32$ square lattice with $U=-4t$ (except in (f)). $%
T=0$. In (f) $\protect\alpha=0.75t$. BCS: uniform superfluid; NG: normal
gas; IN: insulator;}
\label{phase}
\end{figure}

\section{Phase diagram}

In Fig.~(\ref{phase})a-d, we plot the zero-temperature mean-field phase
diagram in the $(\mu ,h)$ plane for $U=-4t$ without and with SO coupling.
Here the parameters correspond to a typical set of parameters $t\sim 2KHz$, $%
U\sim -8KHz$, $\alpha \sim 1.5KHz$ \cite{Gong} in experiments, where $U$ and
$\alpha $ can be respectively tuned through Feshbach resonances~\cite%
{Chin2010RMP} and coherent destructive tunneling methods~\cite%
{Yongping2013SR}.
The phase diagram is symmetric about $\mu =0$ and $\mu =-U/2$, respectively
for the cases without HFC and with HFC due to the particle-hole symmetry.
Since the Zeeman field is along the $x$ direction, we can define the
particle-hole operator, $\mathcal{C}\left(
\begin{array}{c}
\hat{c}_{i\uparrow } \\
\hat{c}_{i\downarrow }%
\end{array}%
\right) \mathcal{C}^{-1}=e^{i\mathrm{\mathbf{\pi }}\cdot \mathrm{\mathbf{R}}%
_{i}}\left(
\begin{array}{c}
\hat{c}_{i\downarrow }^{\dagger } \\
-\hat{c}_{i\uparrow }^{\dagger }%
\end{array}%
\right) $ and $\mathcal{C}\left(
\begin{array}{c}
\hat{c}_{i\uparrow }^{\dagger } \\
\hat{c}_{i\downarrow }^{\dagger }%
\end{array}%
\right) \mathcal{C}^{-1}=e^{-i\mathrm{\mathbf{\pi }}\cdot \mathrm{\mathbf{R}}%
_{i}}\left(
\begin{array}{c}
\hat{c}_{i\downarrow } \\
-\hat{c}_{i\uparrow }%
\end{array}%
\right) $ with $\mathrm{\mathbf{\pi }}=(\pi ,\pi )$ and $\mathcal{C}^{2}=1$.
Under this transformation, $\mathcal{C}H(\mu )\mathcal{C}^{-1}=H(-\mu )$,
leading to the symmetric phases about $\mu =0$ observed in the numerical
calculations. While in the presence of HFC, we have $\mathcal{C}H(\mu )%
\mathcal{C}^{-1}=H(-\mu -U)$ and the phase diagram is now symmetric about $%
\mu =-U/2$.

\begin{figure}[t]
\includegraphics[width=3.5in]{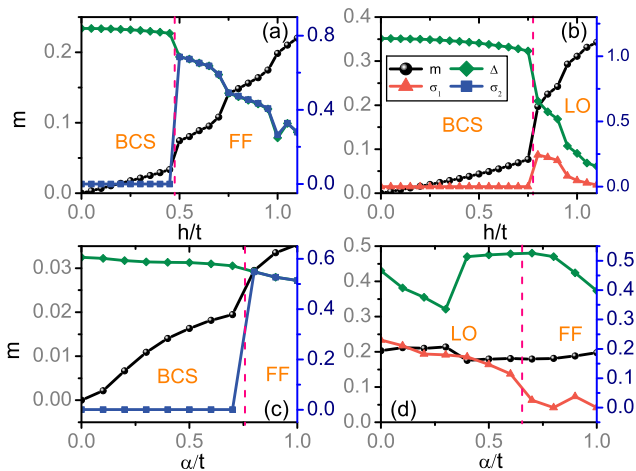}
\caption{(Color online) Change of the average order parameter $\Delta $
(diamond green line), magnetization $m$ (circle black line) defined as $%
m=\sum_{i}\langle \hat{m}_{i}\rangle /N$ with $\hat{m}_{i}=\hat{c}%
_{i\uparrow }^{\dagger }\hat{c}_{i\downarrow }+\hat{c}_{i\downarrow
}^{\dagger }\hat{c}_{i\uparrow }$, amplitude variation $\protect\sigma _{1}$
(triangle red line), and phase variation $\protect\sigma _{2}$ (square blue
line) across the phase transitions. $\protect\alpha =0.75t$ for (a,b). (a) $%
\protect\mu =-2t$; (b) $\protect\mu =-0.4t$; (c) $\protect\mu =-3t$, $h=0.3t$%
; (d) $\protect\mu =-1.5t$, $h=0.9t$. The left vertical axis is for $m$,
while the right axis is for $\Delta $, $\protect\sigma _{1}$, and $\protect%
\sigma _{2}$. The vertical pink lines indicate the phase transition points.
The unit of $\Delta $, $\protect\sigma _{1}$, and $\protect\sigma _{2}$ is $t
$. }
\label{mag}
\end{figure}

To distinguish different phases, we define the order parameter's amplitude
variation $\sigma _{1}=\sqrt{\sum_{i}(|\Delta _{i}|-\overline{|\Delta |}%
)^{2}/N}$ with $\overline{|\Delta |}=\sum_{i}|\Delta _{i}|/N$ and phase
variation $\sigma _{2}=\sqrt{\sum_{i}|\Delta _{i}-\overline{\Delta }|^{2}/N}$
with $\overline{\Delta }=\sum_{i}\Delta _{i}/N$. The normal superfluid phase
is characterized by $\overline{|\Delta |}\neq 0$, $\sigma _{1}=\sigma _{2}=0$%
, the LO phase $\overline{|\Delta |}\neq 0$, $\sigma _{1}\neq 0$, while the
FF phase $\overline{|\Delta |}\neq 0$, $\sigma _{1}=0$, $\sigma _{2}=%
\overline{|\Delta |}$. Generally, a local superfluid order parameter can be
written as $\Delta _{i}=\Delta _{+}\exp \left( iyQ_{y}\right) +\Delta
_{-}\exp \left( -iyQ_{y}+\phi \right) $, where $\phi $ is the relative phase
between two $\pm \mathrm{\mathbf{Q}}$ components. For normal superfluid
phases, $Q_{y}=0$ while for LO and FF phases $Q_{y}\neq 0$. For LO phases,
both $\Delta _{+}$ and $\Delta _{-}$ are nonzero, while for FF phases one of
them is zero. Note that both normal and insulator phases have $|\Delta |=0$,
but the excitations are gapless (gapped) for the normal (insulator) phase.

From Fig.~\ref{phase}a, we see that without SO coupling, there is a large
area of LO phase occupying the region with higher $h$ and no FF phase is
found, which is consistent with the previous report~\cite{Trivedi2010PRL}.
However, in the presence of SO coupling (Fig.~\ref{phase}b), the FF phase
emerges and its existence region is greatly enlarged with the increasing SO
coupling. The region comes from the LO phase, the normal phase, as well as
the BCS phase regions. This implies that the SO coupling enhances $h_{c2}$%
~for the transition to normal phase \cite{Barzykin2002PRL,Partrick2012PRL},
and reduces the $h_{c1}$ from BCS to FF phases. Furthermore, the LO region
is reduced toward $\mu =0$ (half filling). This is a clear competition
between FF and LO phases in the presence of SO coupling, which is more
explicitly shown in Fig.~\ref{phase}e. With increasing SO coupling, the LO
phase region becomes smaller whereas the FF phase region becomes larger.
Fig.~\ref{phase}f shows the effects of interactions on the phase diagram. On
one hand both BCS and FF phases are enhanced with increasing interactions,
on the other hand, larger interactions are capable of inducing the LO phase.

In Fig.~\ref{phase}c, d, we plot the phase diagram in the presence of HFC to
show their effects. There is an insulator region around $\mu =-U/2$ without
SO coupling, which is caused by the nesting of the Fermi surface. With SO
coupling, the LO phase is shrunk in the $\mu $ direction due to the
effective chemical potential shift caused by the Hartree term, while
increased in the $h$ direction, compared with Fig.~\ref{phase}b. The figure
suggests that HFC only quantitatively changes the phase diagram. We also
have confirmed that HFCs have no significant effects on the magnetization
presented in Fig. \ref{mag}. Therefore, we focus on the case without HFC in
the following discussion. 

Generally, traditional $s$-wave BCS pairings do not support finite
magnetization because of equal contributions from both spins. The superfluid
phases that support the coexistence of superfluidity and magnetism
correspond to FFLO phases or Sarma phases~\cite{Sarma1963,Liu2003PRL}, which
are usually gapless. However, in the presence of SO coupling and a Zeeman
field along the $x$ direction, a BCS pairing with gapped excitations also
has a finite magnetization as observed in Figs.~\ref{mag}a-c. Figs.~\ref{mag}%
a and \ref{mag}b respectively display the phase transition from BCS states
to FF states and from BCS states to LO states at certain parameter set
points with a fixed SO coupling as the Zeeman field increases. The
transition is respectively manifested by the dramatic increases of the
variations $\sigma _{2}$ and $\sigma _{1}$. We have also checked the Fourier
transformation of the order parameter $\Delta $ and it shows that apart from
nonzero $Q_{y}$, $\Delta _{+}\neq 0$ and $\Delta _{\_}=0$ for the FF phase
while neither vanishes for the LO phase. The phase variation of FF state can
also be seen from the fact that $\sigma _{2}$ is equal to $\Delta $. Figs. %
\ref{mag}c and \ref{mag}d show the transition from BCS to FF phases and from
LO to FF phases at certain points with a fixed Zeeman field as the SO
coupling $\alpha $ is increased, indicating the growth of FF phases and the
suppression of BCS and LO phases by the SO coupling. Around the transition
point from LO to FF phases, there is no clear change of the magnetization.
The kink of $\Delta $ around $\alpha =0.3$ in Fig.~\ref{mag} d (not at the
phase transition point) is caused by the change of periodicity of $\Delta $
in the LO phase.

\begin{figure}[t]
\includegraphics[width=3.45in]{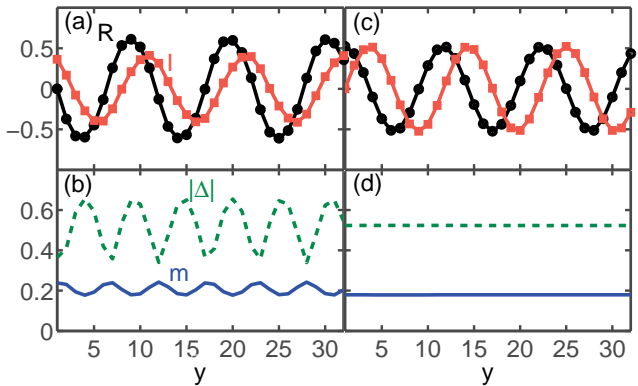}
\caption{(Color online) The order parameter and magnetization in real space
along the $y$ direction. (a)(b) are for the generalized LO phase with $%
\protect\alpha =0.75t$, $\protect\mu =-1.2t$, $h=0.9t$ and (c)(d) are for
the FF phase with $\protect\alpha =0.75t$, $\protect\mu =-1.5t$, $h=0.9t$.
The circle black and square red lines correspond to the real and imaginary
parts of $\Delta $ with the unit $t$.}
\label{real_space}
\end{figure}

\section{Real space signatures}

Previously the LO phase in lattice models is generally characterized by an
inhomogeneous real order parameter and the existence of domain walls at the
node points $\Delta _{i}=0$ that contain the largest magnetization~\cite%
{qinghong2008PRB,Trivedi2010PRL}. These results are reproduced in our
calculation without SO coupling. However, when SO coupling is included, $%
\Delta _{i}$ is no longer real in the LO phase and does not contain nodes,
as clearly seen from Fig.~\ref{real_space}b. The real and imaginary parts of
the order parameter have different phase and amplitude (i.e., $\Delta
_{+}\neq \Delta _{-}$) (Fig.~\ref{real_space}a), indicating the order
parameter has both phase and amplitude variations , which are very different
from traditional FF or LO phases. Such non-zero $\Delta _{i}$ is caused by
the imbalance between the pairings with momentum $\pm \mathrm{\mathbf{Q}}$,
which will be explained in detail in the next section.
The magnetization $m$ also oscillates in space and reaches the maximum at
the minimum of the absolute $|\Delta |$ (See Fig.~\ref{real_space}b). In FF
phase $\Delta _{i}=\Delta _{0}\exp \left( iyQ_{y}\right) $, hence the phase
varies, but the magnitude of the order parameter and the magnetization are
uniform, as shown in Fig.~\ref{real_space}c,d. The Fourier transformation of
the order parameters shows two peaks at $\pm \mathrm{\mathbf{Q}}$ for the LO
phase, but one peak at $\mathrm{\mathbf{Q}}$ for the FF phase, as expected.

\begin{figure}[t]
\includegraphics[width=3.2in]{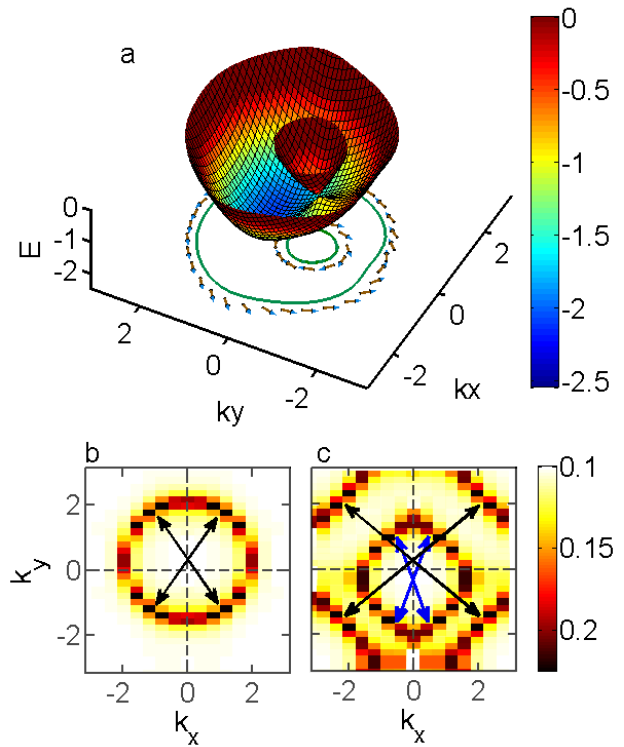}
\caption{(Color online) (a) Single particle band structure in the momentum
space with $\protect\alpha =0.75t$, $h=0.6t$, and $\protect\mu =-3.0t$. The
Fermi surface is plotted on the bottom layer with green line. The small
arrows around the Fermi surface are the spin orientations. (b) shows the
pairing density $|\langle \hat{c}_{\mathrm{\mathbf{k}},-}\hat{c}_{-\mathrm{%
\mathbf{k}}+Q_{y},-}\rangle |^{2}$ with $Q_{y}=3\protect\pi /16a$ and
lattice constant $a$, where $-$ indicates the lower branch. The black double
arrows show the pairing. The parameters are the same as (a). (c) shows the
pairing density $|\langle \hat{c}_{\mathrm{\mathbf{k}},-}\hat{c}_{-\mathrm{%
\mathbf{k}}+Q_{y},-}\rangle |^{2}$, and $|\langle \hat{c}_{\mathrm{\mathbf{k}%
},+}\hat{c}_{-\mathrm{\mathbf{k}}-Q_{y},+}\rangle |^{2}$ for the LO phase
with $\protect\alpha =0.75t$, $h=0.85t$, $\protect\mu =-0.2t$. Here $+$
indicates the higher branch. The black and blue double arrows illustrate the
Cooper pairings with $Q_{y}$ and $-Q_{y}$, respectively. $Q_{y}$ depends on
the deformation of the Fermi surface by the SO coupling and the Zeeman
field. The units of $E$, and $k_{x}$, $k_{y}$ are $t$ and $1/a$ respectively.
}
\label{FF_pairing}
\end{figure}

\section{Pairing mechanism}

It is natural to ask why the combination of SO coupling and an in-plane
Zeeman field is capable of enhancing the parameter region for the FF pairing
while suppressing that for the LO pairing. This can be understood from
different pairing densities of these two phases. We find that in the former,
the pairing mainly occurs around the Fermi surface lying at the lower energy
band in the helicity representation, while in the latter, the pairing occurs
at both energy bands. Fig. \ref{FF_pairing}a illustrates the single particle
band structure at a typical FF phase point with $\alpha =0.75t$, $\mu =-3t$,
and $h=0.6t$. The Fermi surface is plotted at the bottom of the box with
green lines. It is clearly seen that an in-plane Zeeman field with SO
coupling leads to the asymmetric Fermi surface around the origin along the $%
y $ direction, i.e., the lower band (denoted as $-$) has the lowest energy
state located at positive $k_{y}$ while the upper one (denoted as $+$) at
negative $k_{y}$. Cooper pairings with approximately opposite spins happens
mainly around the Fermi surface of the lower band as shown in Fig.~\ref%
{FF_pairing}b due to its higher density of states compared with the upper
one. The deformation of the Fermi surface finally leads to the finite
center-of-mass momentum of Cooper pairs.

As shown in Fig.~\ref{phase}b, when the chemical potential ($\mu<0$) is
increased, the system enters into the BCS superfluid region from the FF
region resulted from the decreased deformation of the Fermi surface.
However, for certain larger Zeeman fields, the same manipulation drives the
system to the LO phase. In this phase, primary contributions to Cooper pairs
are not only from fermions around the Fermi surface of the lower band, but
also these around that of the higher band as the pairing density shows in
Fig.~\ref{FF_pairing}c. Since these pairs from different bands have opposite
momenta, the amplitude of the resulted order parameter has spatial
oscillation structure. In contrast with traditional LO phases, the imbalance
(that is, $\Delta _{+}\neq \Delta _{-}$) of the numbers of pairings with
opposite momenta leads to the absence of nodes, giving the generalized LO
phase. It can be seen that the pairing mechanism is very different from the
case without SO coupling, where pairing happens at different bands. In this
case, if there is a non-zero momentum pairing, there is always another
pairing with the opposite momentum to lower the energy~\cite{Trivedi2010PRL}%
. The consequence is the absence of the FF phase and the presence of the LO
phase with nodes.

\section{Experimental observation}

In experiments, the Rashba SO coupling and an in-plane Zeeman field in a
square lattice can be realized using six lasers that couple two different
hyperfine ground states of atoms \cite{XJLiu2013arXiv}. Since we are mainly
interested in the superfluid phase, only a weak optical lattice is needed.
For a typical set of parameters $t\sim 2KHz$, $U\sim -8KHz$, $\alpha \sim
1.5KHz$ \cite{Gong}, the resulting paring order $\Delta \sim 1KHz\sim 50nK$,
which could be further enhanced by increasing the interaction through a
Feshbach resonance. The critical Zeeman field is generally at the order of
0.5 KHz. 
These parameters show that the FFLO phases should be observable with
reasonably low temperature and realistic experimental setup. In experiments,
the magnetization and the pairing order strength can be observed in the
standard spin-resolved time of flight image. While the finite center-of-mass
momentum of the Cooper pairs may be observed using noise-correlation method
or momentum-resolved radio-frequency spectroscopy~\cite{JZhang2013arXiv}. In
previous spin-imbalanced Fermi gases, the observation of the coexistence of
superfluid and magnetism is generally taken as a signature of the FFLO phase
~\cite{Liao2010Nature}. However, it is no longer true in our case because a
BCS phase also has finite magnetization. In order to observe the FFLO phase,
one should detect the pairing momentum or the magnetization oscillation in
the LO phase directly.

\section{Possible observation in SO coupled superconducting thin films}

Finally, we remark that the same physics may also be observed using SO
coupled s-wave ultra-thin superconducting films subject to an in-plane
Zeeman field, which may be realized using an in-plane magnetic field or a
magnetic semiconductor substrate. Recently, superconductivity in the extreme
2D limit (down to two atomic layers) has been observed in experiments for
many materials~\cite{Qikun2010Nature}. In some of these thin films, such as
Pb, strong Rashba spin-orbit coupling exists and can be tuned through a
variable Schottky barrier~\cite{Slomski2013}. Furthermore, the Hc$_{2}$
critical field for these materials can be extremely large for an in-plane
magnetic field \cite{Xiong2011Natrue}. Such spin-orbit coupled ultra-thin
superconducting films open the door for the possible observation of FFLO
phases, similar as the role of semiconductor-superconductor
nano-heterostructures for the recent search of Majorana fermions~\cite%
{MourikScience2012,DengMTNano2012,DasNature2012}. In Pb experiments, a
typical set of parameters is $t\sim 40$ meV, $\alpha \sim 17meV$. \ However,
the interaction $U$ is generally much weaker, leading to an experimentally
observed s-wave order parameter $\Delta \sim 0.7$ meV (corresponds to the Pb
thin film superconducting transition temperature $T_{c}=6$ K). The required
Zeeman field for the phase transition $h<\Delta $, which requires a magnetic
field $B<6$ T for a small $g$-factor $g=2$, which is below the upper
critical magnetic field $B\sim 8$ T. In the thin films, the FFLO vector $%
Q\propto h_{x}$ is much smaller due to the small deformation of the Fermi
surface by a small Zeeman field, and the resulting order parameter
oscillation period should be much longer. In experiments, the local order
parameter minima in the LO state can accommodate normal quasiparticles,
which lead to nonzero differential conductance that can be detected through
local tunneling measurement. In addition, the Josephson junction between a
FFLO superconductor and a conventional BCS superconductor \cite{KYang2000PRL}
can also be used to detect the FFLO phases.

\textit{Acknowledgement:} We would like to thank Li Mao and Myron Salamon
for helpful discussion. Y. Xu would thank Yongping Zhang for careful and
critical revision. This work is supported by ARO (W911NF-12-1-0334), AFOSR
(FA9550-11-1-0313), and NSF-PHY (1249293).

\end{document}